# Saturation of gas concentration signal of the laser gas sensor

Z.Zh. Zhanabaev[1], A.O. Tileu[1], T.S. Duisebayev[1], D.B. Almen[1*]

[1]Faculty of Physics and Technology, al-Farabi Kazakh National University, Almaty, Kazakhstan

**Correspondence:**
[*]A.O.Tileu
e-mail: tileu.ayan@gmail.com

**Keywords**: **fluctuation-dissipation, correlator, photodiode, methane, carbon tetrachloride, ammonia**

**Abstract**

Nowadays it is possible to determine the type of gas with sufficient accuracy when its concentration is less than $10^{-6}$ (in units of $ppm$) fractions using spectroscopic methods (optical, radio engineering, acoustic). Along with this, the value of permissible concentrations of explosive, toxic, harmful to technology and ecology gases is practically important. Known physical experimental studies indicate only a linear dependence of the response of a laser gas sensor at $ppm \gtrsim 10^3$. The research methods for $ppm \lesssim 10^3$ are based on the processes of combustion, microexplosion, structural and phase transformations and are not always applicable in real practical conditions.

The work is devoted to the analysis of experimentally obtained fluctuations caused by a laser beam in a gas in a photodiode (signal receiver) due to its influence not only at the atomic level, but also on the scale of clusters of nanoparticle molecules. The gas concentration is estimated by the fluctuation-dissipation ratio. It is shown that the signal correlator is saturated to a constant value when the quantum (laser photon energy) and thermal (nanoparticle temperature) factors are comparable with an increase in the concentration of the target gas. The critical values of the saturation concentration are determined by the equality of these two factors.

**Introduction**

Determining the type of different gas and its concentration is an important task of production, medicine, safety, and environmental monitoring. Extensive studies have been carried out using sensors with different measurement principles [1, 2, 3]. Laser absorption spectroscopy systems have been developed, multi-pass sensors with lasers of various wavelengths have been developed to obtain spectral absorption lines and transmit information.

The results of laser sensing are usually presented as a linear dependence of the response on the gas concentration. In practice, it becomes necessary to know the concentration of gas $C_*$, at which the saturation of the sensor response occurs. Thus, in [1] an experimental rule is indicated at $C_* \approx 10^4 \ ppm$ (1%) for methane in a coal mine, sound and light alarms are triggered, and at $C_* \simeq 1,5 \ \%$ – production is turned off.

Saturation can also be observed in the signals of nanofilm sensors [4]. These facts indicate the need for a physical analysis of the concentration attenuation of the signal of a commonly used laser gas sensor.

When the laser beam interacts with gas particles, a sensitive element (photodiode), its intensity decreases. Due to the multiple factors of this process (absorption, reflection, polarization of the beam), the presence of technical noise of the devices, the output signal will fluctuate. Therefore, for the problem under consideration, one can choose a general method of statistical physics – the fluctuation-dissipation theorem (FDT). This method has been used to describe molecular, electromagnetic, and quantum phenomena.

Special experiments [5] have shown the bifurcation effect of thermal noise on microelectromechanical systems, the physical content of which is FDT. The possibility of using the fluctuation-dissipation ratio for modeling the spectrum of a condensed medium is considered in [6].

In recent decades, numerous articles have been published in leading physics journals devoted to expanding the formulation and possible applications of FDT from the point of view of nonequilibrium thermodynamics, the theory of nonlinear response away from equilibrium in shear fluids, bulk media, nanosystems, and glassy media. In a comprehensive list of modern research in this field, only the study [7] stands out, as its subject of analysis is distinct from the others discussed above.

The theoretical foundations of further applications of FDT to the optical phenomena of transmission, absorption, and reflection in media with microscopic structures of various geometries are discussed in [8]. The response of the medium to fluctuations in the external field is described by the Bose-Einstein distribution over the frequency $\omega$ and the field is described by the Greens function, which depends on the coordinates. The results of the technique of calculating the matrices of the electromagnetic field components are presented. The results of the fundamental nature of this work show the universal applicability of FDT to phenomena of a wide spatial and temporal spectral scale.

Following the preceding overview of various studies on laser gas sensors, the primary objective and novelty of this work are formulated. In well-known studies [1], at a low concentration of the target gas, its presence is determined with great accuracy by the characteristic atomic spectral absorption lines of the laser beam. However, as we indicated in the Abstract section, the study of physical phenomena occurring at high gas concentrations is of great practical importance.

With increasing gas concentration, the intensity of the transmitted radiation decreases over distance according to the Beer-Lambert law and at a fixed position of the sensor. At the same time, the proportion of absorption by molecular clots increases, and the power spectrum will be predominantly low-frequency. The intensity of the beam will be comparable to the thermal background. The mutual influence of these two factors is taken into account by the FDT formula. Therefore, it is possible to determine the effect of concentration on the change in the correlation function of fluctuations in the sensor output signal. The comparability of the energy of the average number of quanta of vibrational or rotational modes with the thermal energy of gas molecules will lead to saturation, the nonlinear nature of the change in the signal of the laser sensor.

The experiment will use a time-fluctuating signal of a laser beam transmitted through a layer of gas with different concentrations to a photodiode. At the same time, the conditions of partial coherence of the fluctuation signal are observed, which are necessary to obtain correlation and spectral characteristics.

**Theoretical foundations of the experiment**

The fluctuation-dissipation ratio (FDR) is written in various forms according to the types of characteristic variables. We will use the values of the time fluctuations of the output voltage from the photodiode $v(t_i) \equiv v_i$, the energy of the laser photon $\hbar\omega_0$, the thermal energy of the gas molecules $kT$ ($k$ – the Boltzmann constant). The generally accepted recording of FDT in statistical physics in spectral form by frequency $\omega$ [8]:

$$\langle v_i, v_j \rangle_\omega = \hbar\alpha''(\omega) cth\left(\frac{\hbar\omega}{2kT}\right) = 2\hbar\alpha''(\omega)\left\{\frac{1}{2} + \frac{1}{e^{\hbar\omega/kT}-1}\right\} \quad (1)$$

where $\alpha''(\omega)$ – the imaginary part of generalized susceptibility. The curly bracket expresses the average energy of the oscillator in units of $\hbar\omega$ at temperature $T$. The left part (1) is equal to the Fourier spectrum of the correlation function (correlator) in frequency $\omega$:

$$K_\omega(v(t)) = \left\{\frac{1}{N^2}\sum_{i=1}^{N}\sum_{j=1}^{N} v(t_i)v(t_j)\right\}_\omega \equiv \langle v_i, v_j \rangle_\omega \quad (2)$$

Let us take into account the conditions for completing our task. The concentration of gas $C$ is usually determined in millionths ($ppm$) of the mass or volume of air. Therefore, the correlator of the mixture gas(g)+air(a) $K(v(t); g + a)$ must be attributed to its value for air:

$$K(v(t); g + a)/K(v(t); a) \quad (3)$$

Next, we establish the dependence of expression (3) only on concentration. Therefore, we are looking for the desired dependence relative to a fixed laser frequency $\omega_0$ or a fixed modulation fundamental frequency according to the experimental condition. Then, in the right part of formula (1), the constant value $\hbar\alpha''(\omega_0)$ we can include it in expression (3), which also clearly does not depend on concentration. Let's include a constant expression in the experimentally determined value of expression (3).

$$\hbar\alpha''(\omega_0; g + a)/\hbar\alpha''(\omega_0; a) \quad (4)$$

Thus, we have eliminated the frequency dependence and will proceed with using the original formula (2) without the frequency index $\omega$. This is possible due to the equality of the integral characteristics in time and frequency of oscillations.

After that, in the left part of formula (1), we have a dimensionless expression (3) divided by a dimensionless constant (4). In the right part of formula (1), the number of absorbed coherent photons $cth\left(\frac{\hbar\omega_0}{2kT}\right)$ remains according to the accepted condition $\omega = \omega_0$. This hyperbolic function in formula (1) contains an exponent. The ratio of the close values in formula (3) in comparison with the exponent is calculated more accurately by the natural logarithm:

$$ln\{K(v(t); g + a)/K(v(t); a)\}/K_{max} = cth\left(\frac{\hbar\omega_0}{2kT}\right) \quad (5)$$

The logarithmic multiplier is calculated in a normalized form by dividing by its maximum value modulo $K_{max}$. This takes into account the negative sign in the balance of fluctuation (positive) and dissipation (negative) energies.

The concentration of gas $C$ is clearly not taken into account in formula (5). The concentration value $C_{*0}$, corresponding to saturation of the sensor signal is determined from the equality of fluctuation coherent quantum ($\hbar\omega_0/2$) and dissipation thermal ($kT$) factors:

$$C_{*0} = 2kT/\hbar\omega_0 \tag{6}$$

$C_{*0}$ is determined experimentally at constant values of $\omega_0, T$. In cases of partial coherence, bifurcations of oscillations $C_*$ — the critical value of the onset of saturation differs from $C_{*0}$ (section 4). The hyperbolic function in (5) decreases at $C \to 2C_*$ and remains constant, close to one. To describe the increase in correlations of fluctuations in the right part of formula (5) at $C \lesssim 2C_*$, we use the inverse function $th(x) = cth^{-1}(x)$. Physically, this means that at $C \lesssim 2C_*$ the fluctuating energy of the laser photons is absorbed by the gas, at $C > 2C_*$ it becomes dissipative, comparable to the thermal energy of molecules. Taking into account the influence of the gas concentration, we write formula (5) in a scale-invariant form relative to the maximum of the $K_{max}$ correlator for the case $2C \leq C_*$:

$$ln\{K(v(t,C); g+a)/K(v(t,C); a)\}/K_{max} = th(2C/C_{*0}) \tag{7}$$

**Experimental setup**

The measurements were carried out using the instruments and devices shown in Figure 1. From the GAS BALLOON, through the pressure gauge and VALVE, the gas enters a 20 ml vessel, where an air + gas mixture is formed. The pressure gauge maintains the pressure of the incoming gas equal to atmospheric pressure. A semiconductor LASER with a wavelength of 532 nm is connected to a POWER SUPPLY source with voltage and current indicators. The laser beam passes through the gas+air mixture to the PHOTODIODE PIN. The output signal of the photodiode is fed through the TL081CP operational amplifier to the NIELVISII+ electronics engineering kit. Fluctuation signals from the measuring complex are processed by COMPUTER.

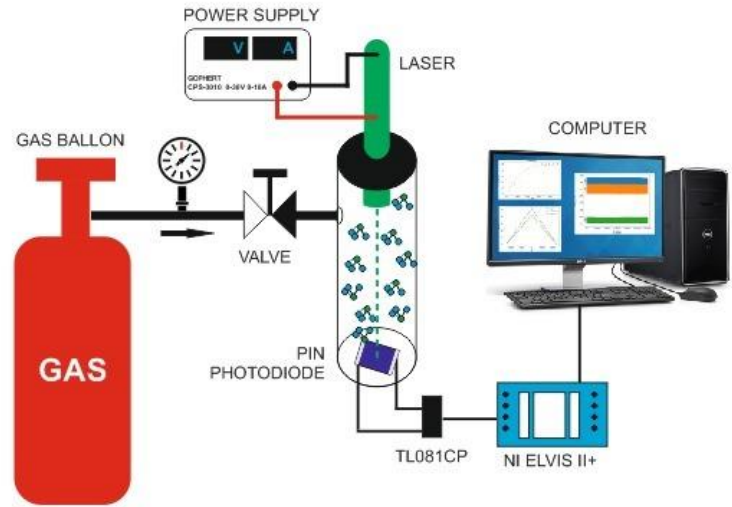

**Figure 1.** Nodes of the laser gas sensor

A Vishay BPW34 silicon P-i-N photodiode was used. The spectral sensitivity line does not coincide with the laser line. This is according to the statement of our task. We are not looking for atomic spectra, but for fluctuations from laser power. Among the available small laser sources, the green laser type SD 303 HUONIE has a higher power of 100 mW than the red laser type FLASH LUK5K 5 mW. In our experiment, the laser output power was controlled by the power supply; The supply current was 60 mA and the voltage was 1.8 V, which corresponds to a power (P=I*U) of 100 mW for a green laser.

**Results of the experiment and discussion**

Stable correlations of the fluctuation signal can be obtained at a degree of coherence of their intensity $\gamma \gtrsim 0,1$, which corresponds to the recommendations adopted in optics for choosing the contrast of partially coherent interference fringes. In the experiment, the required degree of coherence of fluctuations was achieved by the condition

$$\hbar\omega_0 \gtrsim U \qquad (8)$$

where $\hbar\omega_0 = 2,3$ eV – the energy of one laser photon with a wavelength of 532 nm, $U \geq 1\,V$ – the energy of one electron in eV, the stability region of the photodiode characteristic.

Figure 2 shows the time series of partially coherent fluctuations $v(t)$ obtained when a laser beam passes through air (a), through a mixture of air +ammonia at $C = 1200\,ppm$ (b).

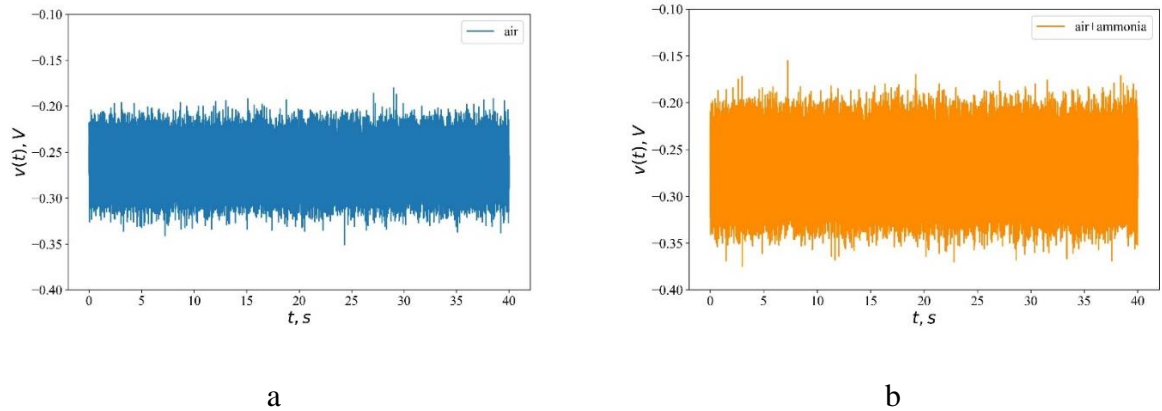

a          b

**Figure 2.** Time series of partially coherent fluctuations $v(t)$:

air, b) air+ammonia, $C = 1200\,ppm$

The degrees of coherence of the interference fringes were determined by the intensity of signal fluctuations

$$\gamma = \frac{[1-(v_{min}/v_{max})^2]}{[1+(v_{min}/v_{max})^2]} \qquad (9)$$

where $v_{min}$, $v_{max}$ – minimum and maximum values of fluctuation voltage, $\gamma \gtrsim 0,2$. We use formula (9) to control the power of laser radiation, which provides a contrasting picture of correlations. For the modulation pattern, $v_{min}/v_{max}$ is used.

The correlation functions $K(v(t))$ of the voltage fluctuations of the sensor $v(t)$, are shown in Figure 3. The dashed approximation lines show the stability of the characteristic for 40 seconds with deviations of 1% (for clean air), 5% (for a mixture of air+gas). During processing, the data was divided into n=10 windows of 4 seconds each. Figure 3 allows us to estimate the total value of the noise intensity under the experimental conditions of fluctuations from air and technical instrument noise. The difference in maximum correlations (variances) is 4%, the standard deviation is 2%. This estimate is comparable to the thermal noise in the purely coherent mode (about 1%) determined by formula (12).

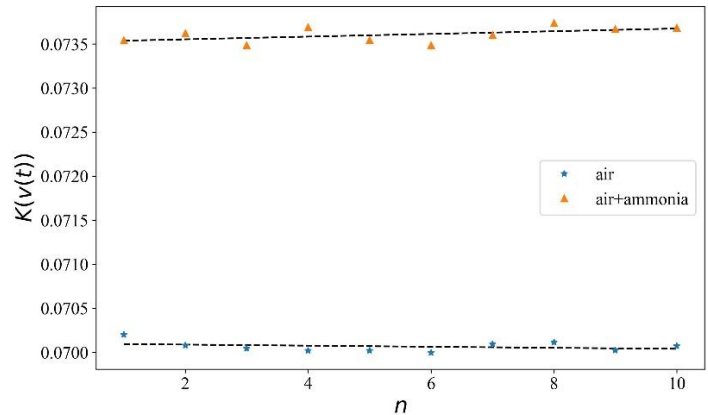

**Figure 3.** Correlation functions of sensor voltage fluctuations $v(t)$ from air (background noise) and from a mixture with gas, corresponding to the data in Figure 2

The relative correlation functions calculated using the left side of formula (5) for gases will be smaller in magnitude than for air.

Figure 4 shows the experimental dependences of the relative values of the $K/K_{max}$ correlation functions on the concentration of $C(ppm)$ gases selected with varying degrees of symmetry of the structure of their molecules (methane $CH_4$, carbon tetrachloride $CCl_4$, ammonia $NH_3$) [9].

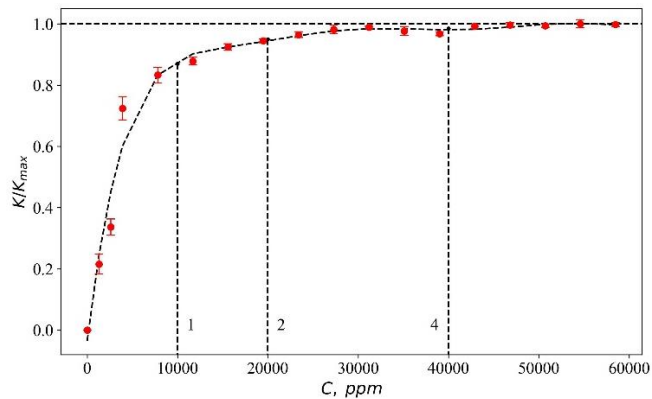

a

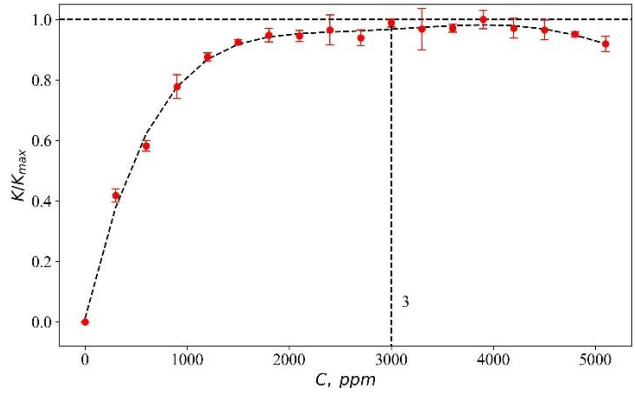

b

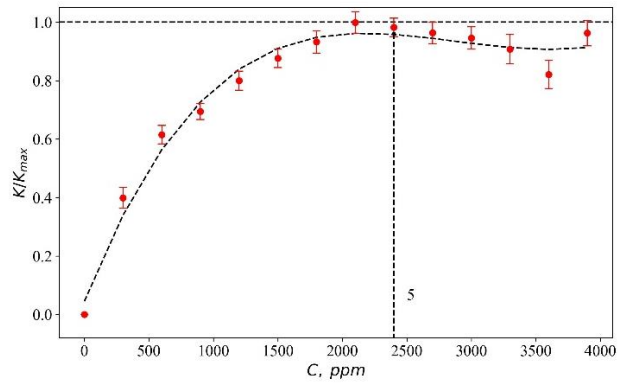

c

**Figure 4**. Change in correlation functions from the concentration of a) methane b) carbon tetrachloride c) ammonia

The ordinate axis shows the values

$$K/K_{max} = ln\{K(v(t); g + a)/K(v(t); a)\}/ K_{max} \quad (10)$$

Experimental values of $K_{max} = 0{,}33\ (CH_4);\ 0{,}11\ (CCl_4);\ 0{,}058\ (NH_3)$. The response of the laser sensor is saturated with an increase in the gas concentration $C \geq C_*$. Transient (bifurcation) modes $C = C_*$ occur near $K \approx K_{max}$. These points in the figures are connected by vertical lines that pass through $C_*(CH_4) = (1,2,4) * 10^4 ppm$, $C_*(CCl_4) = 3 * 10^3 ppm$, $C_*(NH_3) = 2{,}5 * 10^3 ppm$. Such a pronounced concentration saturation of the response of a laser gas sensor can be explained and described. The observed and theoretically described set of (1,2,4) values of $C_*$ is explained separately below.

As the gas concentration increases, the absorption of the laser beam increases, and its intensity becomes comparable to the thermal energy of molecules. The energy of the average number of photons $\langle n \rangle$ according to the formula (1) is equal to

$$E(\langle n \rangle) = \hbar\omega \left(\langle n \rangle + \frac{1}{2}\right) \tag{11}$$

In the symmetric state $\langle n \rangle = 0$ and we have the energy of the coherent state of photons. The equality of the quantum factor at the concentration of photons $C_{*0}$ and the thermal factor is expressed as

$$C_{*0} \hbar\omega_0 = 2kT \tag{12}$$

Under the conditions of our experiment room temperature $T = 300°K$, multiplied by the Boltzmann constant $kT = 10^{-2} eV$, $\lambda_0 = 532$ nm, from $\omega_0 = \frac{2\pi}{\lambda_0} * c$, $\hbar = h/2\pi$ – Planck's constant, $c$ – speed of light, $\hbar\omega_0/2 = 1,15\ eV$ have $C_{*0} = 10^{-2}$ in units of ppm (part per million), $C_{*0}(ppm) = 10^6 * 10^{-2} = 10^4\ ppm$. In methane $(CH_4)$ with a symmetrical arrangement of molecules, saturation of the laser sensor signal occurs at the highest $C_{*0}$. This value $C_{*0}$, equal to 1% of the same order, at which in practice a warning alarm is given in mines [1].

With asymmetry ($\langle n \rangle \neq 0$) of the molecular structure, their vibrations will be partially coherent. Condition (12) will be fulfilled at lower values of $C_*$. This follows from formula (11) if the total energy of the photon is represented in terms of $C_*$, resulting in saturation of the signal. Denoting $\langle n \rangle = C_*$ and the corresponding number of coherent photons $C_*/2$, we have

$$E(C_*) = C_* \hbar\omega \left(1 + \frac{1}{2}\right) = \frac{3}{2} C_* \hbar\omega \tag{13}$$

From a comparison with $E(\langle n \rangle = 0) = C_{*0} \hbar\omega/2$ it follows that $C_* = C_{*0}/3$, indicating that $C_*$ is three times less than for the symmetric case (methane). For example, for gas $CCl_4$ $C_* = C_{*0}/3 = 3,3 * 10^3\ ppm$. The symmetry class $NH_3$ is lower than that of $CH_4$, $CCl_4$. This is due to the additional bonds between the molecules.

In general, the issue boils down to taking into account the correction of the anharmonicity of vibrations and rotations. Well-known publications [10] indicate the possibility of taking into account the $\frac{1}{2} < \alpha < \frac{3}{4}$ correction to formula (11) instead of $\frac{1}{2}$ based on the results of quantum mechanical calculations with various potential barriers for the electron.

We consider it possible to use the correction $\alpha$ of the fluctuation phenomena of the probabilistic information-entropy measure of self-similarity in the range $0,567 \leq \alpha \leq 0,806$, corresponding to the chaotic (to stochastic) regime of the phenomenon [11]. These criteria extend the «golden section» of the dynamic measure (the Fibonacci number is 0.618) to probabilistic phenomena.

Let us explain this statement. From the definition of information $I = -lnP$ through probability $P$ its self-similar value (fixed point $P(I) = I$) is equal to

$$e^{-I_*} = I_*, I_* = 0.567 \tag{14}$$

The average value of information (information entropy $H(I)$) and its self-similar value $I_{*,H}$ are equal:

$$H(I) = -\int_I^\infty e^{-I} I \, dI = (I+1)e^{-I} \qquad (15)$$

$$(I_{*,H} + 1)e^{-I_{*,H}} = I_*, \quad I_{*,H} = 0.806 \qquad (16)$$

Formula (15) takes into account that the probability distribution functions, and its density are equal to $e^{-I}$, which follows from the probability normalization condition. The criteria $I_*, I_{*,H}$ are easily proved by nonlinear (Feigenbaum-type) maps of dynamic chaos. The Fibonacci criterion of 0.618 follows from (15) when taking into account the first term of the exponent expansion at $I < 1$, and at $I \ll 1$ the result (14) will be obtained.

In formula (13), instead of 1/2 we substitute the correction $\alpha = 0{,}806$ (according to formula (16)), and to take into account the concentration we take $\hbar\omega_0 <n> = \hbar\omega_0 C$, $\langle n \rangle = 1$. Comparing $C(NH_3)(1 + 0{,}806) = C(CCl_4)\left(1 + \frac{1}{2}\right)$, we have $C(NH_3) = \frac{3}{2*1.806} C(CCl_4) = 0{,}83 * 3{,}3 * 10^3 = 2{,}5 * 10^3 \, ppm$.

It follows from formula (7) that the graphs of Figure 4 can be constructed in scale-invariant form in the coordinates $th(2C/C_{*0})$ and $C, ppm$ (Figure 5). Such a representation can be useful for establishing possible variations of the parameters $\omega_0, kT, C_*$.

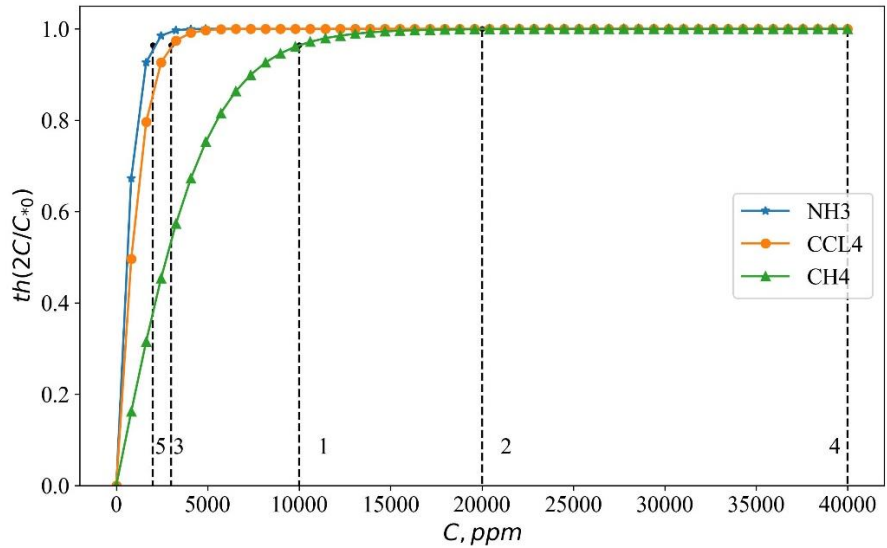

**Figure 5.** Experimental dependence of the dissipative factor in formula (7). $CH_4(1,2,4), CCL_4(3), NH_3(5), C_{*0} = 10^4 ppm$, values (5,3,1,2,4) according to Figure 4

In this regard, we note that due to the fluctuation nature of the laser gas sensor signal under consideration, the measured correlation and dissipation characteristics through hyperbolic functions have harmonic saturation components. This is noticeable from Figures 4. Therefore, the minimum values of $C_*$, are selected, corresponding to the nonlinear behavior of the sensor response.

It is necessary to establish a connection between the minimum $C_*$ and the experimental values of $C_{0_\alpha}$, which differs from $C_{*0}$ for the coherent state of laser photons.

Due to the low volume density compared to the mass density, additional bifurcations with new values of $C_{*_B} > C_*$ are possible in gases. This case can significantly occur in the methane in the form $C_{*_B} > C_{*0}$, because $C_{*0}$ was determined for the coherent States of photons with the symmetric arrangement of atoms. When taking into account anharmonism with the parameter α, we have $C_{*_\alpha} < C_{*0}$.

The essence of the additional bifurcation is as follows. In the presence of a fluctuating effect, the amplitude of the oscillatory process in one period is lower from the equilibrium, in the next it is higher. After a period, after 4 pulses modulo, a bifurcation occurs in the form of a burst. In the equilibrium process as a whole, energy is conserved. This phenomenon in the theory of dynamic chaos is called the period doubling bifurcation. This bifurcation is a universal phenomenon of nature, it is scale-invariant, described by a functional equation [12].

In our problem, due to the accepted condition, the saturation of the signal $\frac{C_* \hbar \omega_0}{2kT} = 1$, at constant $\omega_0$, $T$, $C_* \sim \omega_0^{-1}$. Next, instead of the frequency, it will use $C_{*0}^{-1}$, $C_{*_B}^{-1}$. Scale invariance is manifested at multiple frequencies of $\omega_0$, which is the fundamental frequency, subharmonics $\frac{\omega_0}{2}$, $\frac{\omega_0}{4}$, .... Certainly, in accordance with this in the experiment (Figure 4), the first harmonics can be specified as $C_{*_B} = 2C_{*0}$, $C_{*_B} = 4C_{*0}$, $C_{*0} = 10^4$.

Thus, along with the set of values $C_{*,\alpha}$ according to the anharmonicity parameter (signal nonlinearity), there are also bifurcation values $C_{*_B}$, caused by fluctuations in gas volume. The sensor response – correlator responds to the joint effects $C_{*_\alpha} < C_{*0}$, $C_{*_B} > C_{*0}$. Therefore, in the vicinity of $C \approx C_{*0}$ weaker deviations from linearity of the sensor response relative to its established maximum value are observed when $C_{*_B} > C_{*0}$. This symptom of signal saturation is used in practice. As we have pointed out, the methane in mines prevention and appropriate measures are taken with $C_{*0} = 10^4 \, ppm$ (1%) and $C_{*_B} = 1{,}5 * C_{*0}$.

Let us establish a relationship between $K_{max}$ and the parameter, which takes into account anharmonicity. Taking into account the formula (13), we have

$$C_*(1 + \alpha) = C_{*0}/2 \tag{17}$$

$$\tfrac{1}{2} < \alpha < \tfrac{3}{4} \quad \text{or} \quad 0{,}567 < \alpha < 0{,}806$$

where $C_{*0}$ the saturation concentration of a purely coherent state. The relative correlation of fluctuations is proportional to the number of interacting pairs of particles $\left(K_{max} = \left(\frac{C_\alpha}{C_{*0}}\right)^2\right)$. This ratio and formula (17) summarize experimental and theoretical results.

The above examples of experimental and theoretical estimates show the possibility of a fluctuation-dissipation approach to determining the saturation concentration of a laser gas signal. At the same time, the type of symmetry of gas molecules, fluctuation bifurcations and the condition of comparability of the energies of thermal and quantum perturbations were considered. It is possible to estimate the critical values of the gas concentration without using optical spectral complexes, that is, complex fundamental methods of molecular spectroscopy. Concentration-spectral analysis methods in this direction may be useful to replace specific plasma-photometric, chemiluminescent gas detection methods.

The complexity of the problems and the possibility of continuing in this direction will be indicated by the example of some modern research. The paper [13] presents the results of a theoretical and experimental study of the rotational and vibrational spectra of diamond nanocrystals from meteorites. More than 7000 purely rotational transitions and the first vibrational-rotational states have been detected. Using supercontinuum lasers, ammonia ($NH_3$) and methane ($CH_4$) can be detected in a photonic fiber at a telecommunication wavelength [14]. In [15], Raman scattering of a laser beam with a wavelength of 532 nm was used by a multi-pass ring resonator to detect gases with high sensitivity. Our results were obtained by time correlation of fluctuations at a fixed point. Note that our new physical conclusions about the values of the saturation concentration of the sensor signal, taking into account the nonlinearity due to the anharmonicity of fluctuation bifurcations and the results of repeated measurements correspond within ± 5% of the RMS ratio.

**Conclusion**

In this paper, the possibility of determining the gas concentration values leading to saturation of the response of a laser gas sensor is shown. The correlation functions of fluctuations in the output voltage of a photodiode have been experimentally obtained. The fluctuation-dissipation ratio was used to theoretically describe the experimental results.

It is shown that the nonlinear response of saturation of the laser sensor signal with an increase in gas concentration occurs when the energy of absorbed photons in the gas and the thermal energy of molecules are comparable.

The possibility of using universal functional relations of scale invariance to estimate the values of saturation concentrations of the sensor signal, taking into account the symmetry of the location and the appearance of bifurcations of the oscillation of gas molecules, is shown. This means that the methods of this work are applicable to determine both the gas concentration and the type of target gas.

Knowledge of the onset of the nonlinear sensor operation mode is a practical necessity to prevent disruption of the normal rhythms of life, production, and ecology. In this paper, we have indicated the possibilities of approaching this problem using simple physical methods without using spectral complexes or chemical methods. The experimental technique and measurement methods used by us with the addition of spectral modeling make it possible to create simpler small-sized sensors for the considered problem.